\documentclass[12pt]{article}
\usepackage{a4,epsfig,rotating}

\newcommand{\sqs}{\ensuremath{\sqrt{s}}}
\newcommand{\mh}{\ensuremath{m_{\mathrm H^\pm}}}
\newcommand{\hpm}{\ensuremath{\mathrm H^{\pm}}}

\newcommand{\hh}{\ensuremath{\mathrm{H^+H^-}}}
\newcommand{\hp}{\ensuremath{\mathrm{H^+}}}
\newcommand{\hm}{\ensuremath{\mathrm{H^-}}}
\newcommand{\btn}{\mbox{{\cal B}($\mathrm{H^+\!\to\!\tau^+\nu_\tau}$)}}
\newcommand{\bcs}{\mbox{{\cal B}($\mathrm{H^+\!\to\!c\bar{s}}$)}}
\newcommand{\tntn}{\ensuremath{\tau^+\nu_{\tau}\tau^-\bar{\nu}_{\tau}}}
\newcommand{\tncs}{\ensuremath{\mathrm{c\bar{s}}\tau^-\bar{\nu}_{\tau}}}
\newcommand{\cstn}{\ensuremath{\mathrm{\bar{c}s}\tau^+\nu_{\tau}}}
\newcommand{\cscs}{\ensuremath{\mathrm{c\bar{s}s\bar{c}}}}
\newcommand{\ee}{\ensuremath{\mathrm{e^+e^-}}}
\newcommand{\ww}{\ensuremath{\mathrm{W^+W^-}}}
\newcommand{\qqb}{\ensuremath{\mathrm{q\bar{q}}}}
\newcommand{\qq}{\ensuremath{\mathrm{q\bar{q}^{\prime}}}}

\newcommand{\gevc}{\ensuremath{\mathrm{GeV}/{\it c}}}
\newcommand{\gevcc}{\ensuremath{\mathrm{GeV}/{\it c}^2}}
\newcommand{\invpb}{\ensuremath{\mathrm{pb^{-1}}}}

\font\ninerm=cmr9

\begin{document}

\thispagestyle{empty}

\begin{titlepage}

\begin{picture}(160,1)
\put(2,20){\rm\large EUROPEAN ORGANIZATION FOR NUCLEAR RESEARCH}
\put(115,1){\parbox[t]{45mm}{{\tt CERN EP/2000-nnn}}}
\put(115,-5){\parbox[t]{45mm}{June 29th, 2000}}
\end{picture}

\begin{center}
\vspace{2.1cm}
{\Huge Search for charged Higgs bosons in ${\mathrm e}^+{\mathrm e}^-$ \\
collisions at energies up to \sqs\ = 189~GeV }
\vspace{1.8cm}

The ALEPH Collaboration$^*)$

\end{center}

\vspace{2.2cm}
\begin{abstract}
\vspace{.5cm}
The data collected at centre-of-mass energies of  
188.6~GeV by ALEPH at
LEP, corresponding to an integrated luminosity of
176.2~\invpb, are analysed in a search for pair-produced
charged Higgs bosons~\hpm.
Three analyses are employed to select the \tntn, \tncs{} and \cscs{}
final states.
No evidence for a signal is found.
Upper limits are set on the production cross section as a
function of the branching fraction~\btn{} and of the mass \mh{},
assuming that the sum of the branching ratios is equal to one.
In the framework of a two-Higgs-doublet model, 
charged Higgs bosons with masses below 65.4~\gevcc{} are 
excluded at $95\%$~confidence level 
independently of the decay mode. 
\end{abstract}
\vfill
\centerline{\em (Submitted to Physics Letters B)}
\vskip .2cm
\vskip 1.5cm
\noindent
--------------------------------------------\hfil\break
{\ninerm $^*)$ See next pages for the list of authors}
\end{titlepage}

\clearpage
\pagestyle{empty}
\newpage
\small
%
%
\newlength{\saveparskip}
\newlength{\savetextheight}
\newlength{\savetopmargin}
\newlength{\savetextwidth}
\newlength{\saveoddsidemargin}
\newlength{\savetopsep}
\setlength{\saveparskip}{\parskip}
\setlength{\savetextheight}{\textheight}
\setlength{\savetopmargin}{\topmargin}
\setlength{\savetextwidth}{\textwidth}
\setlength{\saveoddsidemargin}{\oddsidemargin}
\setlength{\savetopsep}{\topsep}
%
%
\setlength{\parskip}{0.0cm}
\setlength{\textheight}{25.0cm}
\setlength{\topmargin}{-1.5cm}
\setlength{\textwidth}{16 cm}
\setlength{\oddsidemargin}{-0.0cm}
\setlength{\topsep}{1mm}
\pretolerance=10000
\centerline{\large\bf The ALEPH Collaboration}
\footnotesize
\vspace{0.5cm}
{\raggedbottom
\begin{sloppypar}
\samepage\noindent
R.~Barate,
D.~Decamp,
P.~Ghez,
C.~Goy,
S.~Jezequel,
J.-P.~Lees,
F.~Martin,
E.~Merle,
\mbox{M.-N.~Minard},
B.~Pietrzyk
\nopagebreak
\begin{center}
\parbox{15.5cm}{\sl\samepage
Laboratoire de Physique des Particules (LAPP), IN$^{2}$P$^{3}$-CNRS,
F-74019 Annecy-le-Vieux Cedex, France}
\end{center}\end{sloppypar}
\vspace{2mm}
\begin{sloppypar}
\noindent
S.~Bravo,
M.P.~Casado,
M.~Chmeissani,
J.M.~Crespo,
E.~Fernandez,
M.~Fernandez-Bosman,
Ll.~Garrido,$^{15}$
E.~Graug\'{e}s,
J.~Lopez,
M.~Martinez,
G.~Merino,
R.~Miquel,
Ll.M.~Mir,
A.~Pacheco,
D.~Paneque,
H.~Ruiz
\nopagebreak
\begin{center}
\parbox{15.5cm}{\sl\samepage
Institut de F\'{i}sica d'Altes Energies, Universitat Aut\`{o}noma
de Barcelona, E-08193 Bellaterra (Barcelona), Spain$^{7}$}
\end{center}\end{sloppypar}
\vspace{2mm}
\begin{sloppypar}
\noindent
A.~Colaleo,
D.~Creanza,
N.~De~Filippis,
M.~de~Palma,
G.~Iaselli,
G.~Maggi,
M.~Maggi,
S.~Nuzzo,
A.~Ranieri,
G.~Raso,
F.~Ruggieri,
G.~Selvaggi,
L.~Silvestris,
P.~Tempesta,
A.~Tricomi,$^{3}$
G.~Zito
\nopagebreak
\begin{center}
\parbox{15.5cm}{\sl\samepage
Dipartimento di Fisica, INFN Sezione di Bari, I-70126 Bari, Italy}
\end{center}\end{sloppypar}
\vspace{2mm}
\begin{sloppypar}
\noindent
X.~Huang,
J.~Lin,
Q. Ouyang,
T.~Wang,
Y.~Xie,
R.~Xu,
S.~Xue,
J.~Zhang,
L.~Zhang,
W.~Zhao
\nopagebreak
\begin{center}
\parbox{15.5cm}{\sl\samepage
Institute of High Energy Physics, Academia Sinica, Beijing, The People's
Republic of China$^{8}$}
\end{center}\end{sloppypar}
\vspace{2mm}
\begin{sloppypar}
\noindent
D.~Abbaneo,
G.~Boix,$^{6}$
O.~Buchm\"uller,
M.~Cattaneo,
F.~Cerutti,
G.~Dissertori,
H.~Drevermann,
R.W.~Forty,
M.~Frank,
F.~Gianotti,
T.C.~Greening,
A.W.~Halley,
J.B.~Hansen,
J.~Harvey,
P.~Janot,
B.~Jost,
M.~Kado,
V.~Lemaitre,
P.~Maley,
P.~Mato,
A.~Minten,
A.~Moutoussi,
F.~Ranjard,
L.~Rolandi,
D.~Schlatter,
M.~Schmitt,$^{20}$
O.~Schneider,$^{2}$
P.~Spagnolo,
W.~Tejessy,
F.~Teubert,
E.~Tournefier,
A.~Valassi,
J.J.~Ward,
A.E.~Wright
\nopagebreak
\begin{center}
\parbox{15.5cm}{\sl\samepage
European Laboratory for Particle Physics (CERN), CH-1211 Geneva 23,
Switzerland}
\end{center}\end{sloppypar}
\vspace{2mm}
\begin{sloppypar}
\noindent
Z.~Ajaltouni,
F.~Badaud,
G.~Chazelle,
O.~Deschamps,
S.~Dessagne,
A.~Falvard,
P.~Gay,
C.~Guicheney,
P.~Henrard,
J.~Jousset,
B.~Michel,
S.~Monteil,
\mbox{J-C.~Montret},
D.~Pallin,
J.M.~Pascolo,
P.~Perret,
F.~Podlyski
\nopagebreak
\begin{center}
\parbox{15.5cm}{\sl\samepage
Laboratoire de Physique Corpusculaire, Universit\'e Blaise Pascal,
IN$^{2}$P$^{3}$-CNRS, Clermont-Ferrand, F-63177 Aubi\`{e}re, France}
\end{center}\end{sloppypar}
\vspace{2mm}
\begin{sloppypar}
\noindent
J.D.~Hansen,
J.R.~Hansen,
P.H.~Hansen,$^{1}$
B.S.~Nilsson,
A.~W\"a\"an\"anen
\nopagebreak
\begin{center}
\parbox{15.5cm}{\sl\samepage
Niels Bohr Institute, 2100 Copenhagen, DK-Denmark$^{9}$}
\end{center}\end{sloppypar}
\vspace{2mm}
\begin{sloppypar}
\noindent
G.~Daskalakis,
A.~Kyriakis,
C.~Markou,
E.~Simopoulou,
A.~Vayaki
\nopagebreak
\begin{center}
\parbox{15.5cm}{\sl\samepage
Nuclear Research Center Demokritos (NRCD), GR-15310 Attiki, Greece}
\end{center}\end{sloppypar}
\vspace{2mm}
\begin{sloppypar}
\noindent
A.~Blondel,$^{12}$
\mbox{J.-C.~Brient},
F.~Machefert,
A.~Roug\'{e},
M.~Swynghedauw,
R.~Tanaka
\linebreak
H.~Videau
\nopagebreak
\begin{center}
\parbox{15.5cm}{\sl\samepage
Laboratoire de Physique Nucl\'eaire et des Hautes Energies, Ecole
Polytechnique, IN$^{2}$P$^{3}$-CNRS, \mbox{F-91128} Palaiseau Cedex, France}
\end{center}\end{sloppypar}
\vspace{2mm}
\begin{sloppypar}
\noindent
E.~Focardi,
G.~Parrini,
K.~Zachariadou
\nopagebreak
\begin{center}
\parbox{15.5cm}{\sl\samepage
Dipartimento di Fisica, Universit\`a di Firenze, INFN Sezione di Firenze,
I-50125 Firenze, Italy}
\end{center}\end{sloppypar}
\vspace{2mm}
\begin{sloppypar}
\noindent
A.~Antonelli,
M.~Antonelli,
G.~Bencivenni,
G.~Bologna,$^{4}$
F.~Bossi,
P.~Campana,
G.~Capon,
V.~Chiarella,
P.~Laurelli,
G.~Mannocchi,$^{5}$
F.~Murtas,
G.P.~Murtas,
L.~Passalacqua,
M.~Pepe-Altarelli
\nopagebreak
\begin{center}
\parbox{15.5cm}{\sl\samepage
Laboratori Nazionali dell'INFN (LNF-INFN), I-00044 Frascati, Italy}
\end{center}\end{sloppypar}
\vspace{2mm}
\begin{sloppypar}
\noindent
M.~Chalmers,
J.~Kennedy,
J.G.~Lynch,
P.~Negus,
V.~O'Shea,
B.~Raeven,
D.~Smith,
P.~Teixeira-Dias,
A.S.~Thompson
\nopagebreak
\begin{center}
\parbox{15.5cm}{\sl\samepage
Department of Physics and Astronomy, University of Glasgow, Glasgow G12
8QQ,United Kingdom$^{10}$}
\end{center}\end{sloppypar}
\begin{sloppypar}
\noindent
R.~Cavanaugh,
S.~Dhamotharan,
C.~Geweniger,$^{1}$
P.~Hanke,
V.~Hepp,
E.E.~Kluge,
G.~Leibenguth,
A.~Putzer,
K.~Tittel,
S.~Werner,$^{19}$
M.~Wunsch$^{19}$
\nopagebreak
\begin{center}
\parbox{15.5cm}{\sl\samepage
Kirchhoff-Institut f\"ur Physik, Universit\"at Heidelberg, D-69120
Heidelberg, Germany$^{16}$}
\end{center}\end{sloppypar}
\vspace{2mm}
\begin{sloppypar}
\noindent
R.~Beuselinck,
D.M.~Binnie,
W.~Cameron,
G.~Davies,
P.J.~Dornan,
M.~Girone,
N.~Marinelli,
J.~Nowell,
H.~Przysiezniak,$^{1}$
J.K.~Sedgbeer,
J.C.~Thompson,$^{14}$
E.~Thomson,$^{23}$
R.~White
\nopagebreak
\begin{center}
\parbox{15.5cm}{\sl\samepage
Department of Physics, Imperial College, London SW7 2BZ,
United Kingdom$^{10}$}
\end{center}\end{sloppypar}
\vspace{2mm}
\begin{sloppypar}
\noindent
V.M.~Ghete,
P.~Girtler,
E.~Kneringer,
D.~Kuhn,
G.~Rudolph
\nopagebreak
\begin{center}
\parbox{15.5cm}{\sl\samepage
Institut f\"ur Experimentalphysik, Universit\"at Innsbruck, A-6020
Innsbruck, Austria$^{18}$}
\end{center}\end{sloppypar}
\vspace{2mm}
\begin{sloppypar}
\noindent
C.K.~Bowdery,
P.G.~Buck,
D.P.~Clarke,
G.~Ellis,
A.J.~Finch,
F.~Foster,
G.~Hughes,
R.W.L.~Jones,
N.A.~Robertson,
M.~Smizanska
\nopagebreak
\begin{center}
\parbox{15.5cm}{\sl\samepage
Department of Physics, University of Lancaster, Lancaster LA1 4YB,
United Kingdom$^{10}$}
\end{center}\end{sloppypar}
\vspace{2mm}
\begin{sloppypar}
\noindent
I.~Giehl,
F.~H\"olldorfer,
K.~Jakobs,
K.~Kleinknecht,
M.~Kr\"ocker,
A.-S.~M\"uller,
H.-A.~N\"urnberger,
G.~Quast,$^{1}$
B.~Renk,
E.~Rohne,
H.-G.~Sander,
S.~Schmeling,
H.~Wachsmuth,
C.~Zeitnitz,
T.~Ziegler
\nopagebreak
\begin{center}
\parbox{15.5cm}{\sl\samepage
Institut f\"ur Physik, Universit\"at Mainz, D-55099 Mainz, Germany$^{16}$}
\end{center}\end{sloppypar}
\vspace{2mm}
\begin{sloppypar}
\noindent
A.~Bonissent,
J.~Carr,
P.~Coyle,
C.~Curtil,
A.~Ealet,
D.~Fouchez,
O.~Leroy,
T.~Kachelhoffer,
P.~Payre,
D.~Rousseau,
A.~Tilquin
\nopagebreak
\begin{center}
\parbox{15.5cm}{\sl\samepage
Centre de Physique des Particules de Marseille, Univ M\'editerran\'ee,
IN$^{2}$P$^{3}$-CNRS, F-13288 Marseille, France}
\end{center}\end{sloppypar}
\vspace{2mm}
\begin{sloppypar}
\noindent
M.~Aleppo,
S.~Gilardoni,
F.~Ragusa
\nopagebreak
\begin{center}
\parbox{15.5cm}{\sl\samepage
Dipartimento di Fisica, Universit\`a di Milano e INFN Sezione di
Milano, I-20133 Milano, Italy.}
\end{center}\end{sloppypar}
\vspace{2mm}
\begin{sloppypar}
\noindent
H.~Dietl,
G.~Ganis,
A.~Heister,
K.~H\"uttmann,
G.~L\"utjens,
C.~Mannert,
W.~M\"anner,
\mbox{H.-G.~Moser},
S.~Schael,
R.~Settles,$^{1}$
H.~Stenzel,
W.~Wiedenmann,
G.~Wolf
\nopagebreak
\begin{center}
\parbox{15.5cm}{\sl\samepage
Max-Planck-Institut f\"ur Physik, Werner-Heisenberg-Institut,
D-80805 M\"unchen, Germany\footnotemark[16]}
\end{center}\end{sloppypar}
\vspace{2mm}
\begin{sloppypar}
\noindent
P.~Azzurri,
J.~Boucrot,$^{1}$
O.~Callot,
M.~Davier,
L.~Duflot,
\mbox{J.-F.~Grivaz},
Ph.~Heusse,
A.~Jacholkowska,$^{1}$
L.~Serin,
\mbox{J.-J.~Veillet},
I.~Videau,$^{1}$
J.-B.~de~Vivie~de~R\'egie,
D.~Zerwas
\nopagebreak
\begin{center}
\parbox{15.5cm}{\sl\samepage
Laboratoire de l'Acc\'el\'erateur Lin\'eaire, Universit\'e de Paris-Sud,
IN$^{2}$P$^{3}$-CNRS, F-91898 Orsay Cedex, France}
\end{center}\end{sloppypar}
\vspace{2mm}
\begin{sloppypar}
\noindent
G.~Bagliesi,
T.~Boccali,
G.~Calderini,
V.~Ciulli,
L.~Fo\`a,
A.~Giammanco,
A.~Giassi,
F.~Ligabue,
A.~Messineo,
F.~Palla,$^{1}$
G.~Rizzo,
G.~Sanguinetti,
A.~Sciab\`a,
G.~Sguazzoni,
R.~Tenchini,$^{1}$
A.~Venturi,
P.G.~Verdini
\samepage
\begin{center}
\parbox{15.5cm}{\sl\samepage
Dipartimento di Fisica dell'Universit\`a, INFN Sezione di Pisa,
e Scuola Normale Superiore, I-56010 Pisa, Italy}
\end{center}\end{sloppypar}
\vspace{2mm}
\begin{sloppypar}
\noindent
G.A.~Blair,
J.~Coles,
G.~Cowan,
M.G.~Green,
D.E.~Hutchcroft,
L.T.~Jones,
T.~Medcalf,
J.A.~Strong,
\mbox{J.H.~von~Wimmersperg-Toeller} 
\nopagebreak
\begin{center}
\parbox{15.5cm}{\sl\samepage
Department of Physics, Royal Holloway \& Bedford New College,
University of London, Surrey TW20 OEX, United Kingdom$^{10}$}
\end{center}\end{sloppypar}
\vspace{2mm}
\begin{sloppypar}
\noindent
R.W.~Clifft,
T.R.~Edgecock,
P.R.~Norton,
I.R.~Tomalin
\nopagebreak
\begin{center}
\parbox{15.5cm}{\sl\samepage
Particle Physics Dept., Rutherford Appleton Laboratory,
Chilton, Didcot, Oxon OX11 OQX, United Kingdom$^{10}$}
\end{center}\end{sloppypar}
\vspace{2mm}
\begin{sloppypar}
\noindent
\mbox{B.~Bloch-Devaux},
D.~Boumediene,
P.~Colas,
B.~Fabbro,
G.~Fa\"{\i}f,
E.~Lan\c{c}on,
\mbox{M.-C.~Lemaire},
E.~Locci,
P.~Perez,
J.~Rander,
\mbox{J.-F.~Renardy},
A.~Rosowsky,
P.~Seager,$^{13}$
A.~Trabelsi,$^{21}$
B.~Tuchming,
B.~Vallage
\nopagebreak
\begin{center}
\parbox{15.5cm}{\sl\samepage
CEA, DAPNIA/Service de Physique des Particules,
CE-Saclay, F-91191 Gif-sur-Yvette Cedex, France$^{17}$}
\end{center}\end{sloppypar}
\vspace{2mm}
\begin{sloppypar}
\noindent
S.N.~Black,
J.H.~Dann,
C.~Loomis,
H.Y.~Kim,
N.~Konstantinidis,
A.M.~Litke,
M.A. McNeil,
G.~Taylor
\nopagebreak
\begin{center}
\parbox{15.5cm}{\sl\samepage
Institute for Particle Physics, University of California at
Santa Cruz, Santa Cruz, CA 95064, USA$^{22}$}
\end{center}\end{sloppypar}
\vspace{2mm}
\begin{sloppypar}
\noindent
C.N.~Booth,
S.~Cartwright,
F.~Combley,
P.N.~Hodgson,
M.~Lehto,
L.F.~Thompson
\nopagebreak
\begin{center}
\parbox{15.5cm}{\sl\samepage
Department of Physics, University of Sheffield, Sheffield S3 7RH,
United Kingdom$^{10}$}
\end{center}\end{sloppypar}
\vspace{2mm}
\begin{sloppypar}
\noindent
K.~Affholderbach,
A.~B\"ohrer,
S.~Brandt,
C.~Grupen,$^{1}$
J.~Hess,
A.~Misiejuk,
G.~Prange,
U.~Sieler
\nopagebreak
\begin{center}
\parbox{15.5cm}{\sl\samepage
Fachbereich Physik, Universit\"at Siegen, D-57068 Siegen, Germany$^{16}$}
\end{center}\end{sloppypar}
\vspace{2mm}
\begin{sloppypar}
\noindent
C.~Borean,
G.~Giannini,
B.~Gobbo
\nopagebreak
\begin{center}
\parbox{15.5cm}{\sl\samepage
Dipartimento di Fisica, Universit\`a di Trieste e INFN Sezione di Trieste,
I-34127 Trieste, Italy}
\end{center}\end{sloppypar}
\vspace{2mm}
\begin{sloppypar}
\noindent
H.~He,
J.~Putz,
J.~Rothberg,
S.~Wasserbaech
\nopagebreak
\begin{center}
\parbox{15.5cm}{\sl\samepage
Experimental Elementary Particle Physics, University of Washington, Seattle,
WA 98195 U.S.A.}
\end{center}\end{sloppypar}
\vspace{2mm}
\begin{sloppypar}
\noindent
S.R.~Armstrong,
K.~Cranmer,
P.~Elmer,
D.P.S.~Ferguson,
Y.~Gao,
S.~Gonz\'{a}lez,
O.J.~Hayes,
H.~Hu,
S.~Jin,
J.~Kile,
P.A.~McNamara III,
J.~Nielsen,
W.~Orejudos,
Y.B.~Pan,
Y.~Saadi,
I.J.~Scott,
J.~Walsh,
J.~Wu,
Sau~Lan~Wu,
X.~Wu,
G.~Zobernig
\nopagebreak
\begin{center}
\parbox{15.5cm}{\sl\samepage
Department of Physics, University of Wisconsin, Madison, WI 53706,
USA$^{11}$}
\end{center}\end{sloppypar}
}
\footnotetext[1]{Also at CERN, 1211 Geneva 23, Switzerland.}
\footnotetext[2]{Now at Universit\'e de Lausanne, 1015 Lausanne, Switzerland.}
\footnotetext[3]{Also at Dipartimento di Fisica di Catania and INFN Sezione di
 Catania, 95129 Catania, Italy.}
\footnotetext[4]{Also Istituto di Fisica Generale, Universit\`{a} di
Torino, 10125 Torino, Italy.}
\footnotetext[5]{Also Istituto di Cosmo-Geofisica del C.N.R., Torino,
Italy.}
\footnotetext[6]{Supported by the Commission of the European Communities,
contract ERBFMBICT982894.}
\footnotetext[7]{Supported by CICYT, Spain.}
\footnotetext[8]{Supported by the National Science Foundation of China.}
\footnotetext[9]{Supported by the Danish Natural Science Research Council.}
\footnotetext[10]{Supported by the UK Particle Physics and Astronomy Research
Council.}
\footnotetext[11]{Supported by the US Department of Energy, grant
DE-FG0295-ER40896.}
\footnotetext[12]{Now at Departement de Physique Corpusculaire, Universit\'e de
Gen\`eve, 1211 Gen\`eve 4, Switzerland.}
\footnotetext[13]{Supported by the Commission of the European Communities,
contract ERBFMBICT982874.}
\footnotetext[14]{Also at Rutherford Appleton Laboratory, Chilton, Didcot, UK.}
\footnotetext[15]{Permanent address: Universitat de Barcelona, 08208 Barcelona,
Spain.}
\footnotetext[16]{Supported by the Bundesministerium f\"ur Bildung,
Wissenschaft, Forschung und Technologie, Germany.}
\footnotetext[17]{Supported by the Direction des Sciences de la
Mati\`ere, C.E.A.}
\footnotetext[18]{Supported by the Austrian Ministry for Science and Transport.}
\footnotetext[19]{Now at SAP AG, 69185 Walldorf, Germany}
\footnotetext[20]{Now at Harvard University, Cambridge, MA 02138, U.S.A.}
\footnotetext[21]{Now at D\'epartement de Physique, Facult\'e des Sciences de Tunis, 1060 Le Belv\'ed\`ere, Tunisia.}
\footnotetext[22]{Supported by the US Department of Energy,
grant DE-FG03-92ER40689.}
\footnotetext[23]{Now at Department of Physics, Ohio State University, Columbus, OH 43210-1106, U.S.A.}
%
\setlength{\parskip}{\saveparskip}
\setlength{\textheight}{\savetextheight}
\setlength{\topmargin}{\savetopmargin}
\setlength{\textwidth}{\savetextwidth}
\setlength{\oddsidemargin}{\saveoddsidemargin}
\setlength{\topsep}{\savetopsep}
\normalsize
\newpage
\pagestyle{plain}
\setcounter{page}{1}

\pagestyle{plain}
\pagenumbering{arabic}

\section{Introduction}

In the Standard Model, particle masses are generated 
via the Higgs mechanism 
implemented using one doublet  of complex scalar fields.
In this process one physical state remains in the spectrum, 
known as the Standard Model Higgs boson. 
The most important phenomenological consequence of an extended 
Higgs sector is the appearance of additional spin-0 states, both 
neutral and charged. For example, with the addition of one more 
doublet, five physical states remain after the spontaneous breaking 
of the $\mathrm{SU(2)_L\times U(1)_Y}$ symmetry: 
three neutral and a pair of charged Higgs bosons.
Among the possible extensions of the Higgs sector, those obtained by adding 
more doublets are preferred because they naturally lead at tree 
level to $\mathrm{M_W\simeq M_Z \cos\theta_W}$, 
a relation very well verified by experiment. 

The ALEPH data collected at centre-of-mass energies up to 184 GeV
have been used in Refs.~\cite{HCH172,HCH183} to search for pair production of  
charged Higgs bosons predicted in models with two Higgs doublets. 
The negative result of the search was translated into a  lower 
limit on the $\hpm$ mass $\mh$ of 59 GeV/c$^2$ at 95\% confidence level. 
In this paper an update of the search based on  
the data collected at $\sqrt{s}= 188.6~\mathrm{GeV}$ (hereafter referred 
to as the 189 GeV data) is presented. 
The theoretical framework and underlying assumptions are the same as detailed 
in Refs.~\cite{HCH172,HCH183}. The $\hp$ is assumed to decay 
predominantly into $\mathrm{c\bar{s}}$ or ${\mathrm \tau^+\nu_\tau}$
final states (and respective charge conjugates for the $\hm$). Other decay 
modes are not considered and~$\btn+\bcs=1$~is assumed, but
the analysis is equally sensitive to other hadronic decay modes.
As a consequence, $\hh$ pair production leads to three 
final states (\cscs{}, \tncs{}/\cstn{} and \tntn) for which separate searches 
are performed.

This letter is organized as follows.
After a brief description of the ALEPH detector in
Section~2, the event selections are described in Section~3. The results and
the conclusions are given in Sections~4~and~5.

\section{The ALEPH detector}
Only a brief description of the ALEPH subdetectors relevant for this 
analysis are given here. A more comprehensive description of the detector 
components is given in Ref.~\cite{bib:detectorpaper} and of the reconstruction 
algorithms in Ref.~\cite{bib:performancepaper}.

The trajectories of charged particles are measured with a 
silicon vertex detector, a cylindrical drift chamber, and a large time
projection chamber (TPC). 
These are immersed in a 1.5~T axial field provided
by a superconducting solenoidal coil. This system yields a resolution of 
$\delta p_T / p_T$ = $6 \times 10^{-4} p_T \oplus 0.005$ ($p_T$ in \gevc{}).
Hereafter, charged particle tracks
reconstructed with at least four hits in the TPC,
and originating from within a cylinder of length~20~cm and radius~2~cm
coaxial with the beam and centred at the nominal collision point,
are referred to as {\it good tracks}.

The electromagnetic calorimeter, placed between the tracking system and the 
coil, is a highly segmented sampling calorimeter which is used to identify 
electrons and photons and to measure their energies. It has a total thickness
of 22 radiation lengths at normal incidence and provides a relative energy
resolution of $0.18 / \sqrt{E} + 0.009$ ($E$ in GeV). 
The luminosity monitors extend the calorimetric coverage
down to 34~mrad from the beam axis.

Muons are identified by their penetration in the hadron calorimeter, 
a 1.2~m thick iron yoke instrumented with 23 layers of streamer tubes, 
together with two surrounding layers of muon chambers.
 The hadron calorimeter also provides a 
measurement of the energy of charged and neutral hadrons with a relative 
resolution of $85 \% / \sqrt{E}$ ($E$ in GeV).

The calorimetry and tracking information are combined
in an energy flow algorithm, classifying a set of
energy flow ``particles'' as photons, neutral
hadrons and charged particles.  
From these objects, jets are reconstructed with an energy resolution of
$( 0.60 \sqrt{E} + 0.60 ) \times (1 + \cos^2 \theta)$ where $E$ in GeV and 
$\theta$ are the jet energy and polar angle, respectively.

\section{Analysis}

To ensure good potential for discovery, independent of the branching
fraction~\btn{}, three selections are defined for the topologies \tntn, 
\tncs{}/\cstn{} (hereafter referred to as \tncs{}) and \cscs.
The most relevant selection criteria 
are chosen to achieve the best expected confidence level 
for exclusion of a mass hypothesis of 70~\gevcc{}. 
Each selection is optimised individually 
with the most optimistic~\btn{} in each case, {\it i.e.}, $0\%$, $100\%$ and 
$50\%$ for the \cscs{}, \tntn{} and \tncs{} channels, respectively.

\subsection{Monte Carlo samples}

Fully simulated Monte Carlo event samples reconstructed with the same program
as the data have been used for background estimates, design of
selections and cut optimization.
Samples of all background sources corresponding to at least
20 times the collected luminosity were generated.
The most important background sources are $\ee\to\tau^+\tau^-$, \qqb{},
four-fermion processes and two-photon collisions, simulated with the
KORALZ~\cite{WAS1}, PYTHIA~\cite{PYTHIA},
KORALW~\cite{KORALW} and PHOT02~\cite{PHOT} generators.

The signal Monte Carlo events were generated using the
HZHA~\cite{JANOT} generator.
Samples of at least 2000 signal events were simulated for each of the
various final states for
charged Higgs boson masses between 50 and 75~\gevcc.

\begin{boldmath}
\subsection{The \tntn{} final state}
\end{boldmath}
The final state produced by leptonic decays of both charged Higgs
bosons consists of two acoplanar $\tau$'s and missing energy carried away
by the neutrinos. As this topology is identical to that expected from 
stau pair production with massless neutralinos,
the ``Large $\Delta M$'' selection described in Ref.~\cite{slepton189} 
is used here to search for charged Higgs bosons in this channel.
Efficiencies to select events from $\hh\to\tntn$ are of the order of $35\%$, 
as shown in Table~\ref{effic}
for a representative set of charged Higgs boson masses. 
The total expected background amounts to 15.5 events, mainly
consisting of irreducible background from $\ww\to\tntn$.
In the data, 20 events are selected, in agreement with 
the expectation. The systematic uncertainty on the number of expected 
signal events is estimated to be $3.0\%$, 
dominated by the effect of limited Monte Carlo statistics ($2.7\%$) 
and uncertainties on the cross-section
for charged Higgs production ($1.0\%$). 
The systematic error on the background is estimated to be $8\%$. This 
is dominated by the effect of limited Monte Carlo statistics ($4\%$),
uncertainties on the cross section for W pair production ($2\%$), and 
uncertainities on the cross section for two-photon processes ($7\%$). The
systematic error on the luminosity is estimated to be 0.5\%.

\begin{table}[t]
 \caption{Selection efficiencies $\epsilon$ (in \%) as a function
of the charged Higgs boson mass $\mh$.}
\begin{center}
 \begin{tabular}{|l|c|c|c|c|c|c|}
\hline
  $\mh$ (\gevcc{}) & 50   & 55    & 60   & 65    & 70   & 75 \\
\hline
\hline
$\epsilon \, (\tntn)$  & 34 & 35 & 38 & 35 & 40 & 41 \\
\hline
$\epsilon \, (\tncs)$  & 35 & 37 & 37 & 35 & 29 & 20 \\
\hline
$\epsilon \, (\cscs)$  & 48 & 48 & 49 & 49 & 48 & 45 \\
\hline

 \end{tabular}
\label{effic}
\end{center}
\end{table}

\begin{boldmath}
\subsection{The \tncs{} final state}
\end{boldmath}

The mixed final state \tncs{} is characterised by two jets originating 
from the hadronic decay of one of the charged Higgs bosons
and a thin jet plus missing energy due to the neutrinos
from the subsequent decay of the charged Higgs and of the $\tau$. 

As a first step in the analysis the thrust of the event is required to be 
less than 0.96 and the total number of good
charged tracks greater than 7. To reduce background from two-photon 
processes and the contribution of beam related background
which is not simulated, 
the energy deposited in a $12^{\circ}$ cone around the 
beam axis is required to be less than $2.5\%$ of the centre-of-mass energy.  
Background from $\rm{e}^+ \rm{e}^- \rightarrow \qqb (\gamma)$ events is reduced
by demanding that the polar angle $\theta_{\rm miss}$ of the missing momentum 
vector point away from the beam axis such that 
${|} \rm{cos}\theta_{miss} {|} 
< 0.9$. To reduce the background from $\ww \rightarrow \qq 
\ell \overline{\nu}_{{\ell}}$ where $\ell$ corresponds to an electron 
or muon, it is required that the events contain no identified lepton
with a momentum greater than $10\%$ of the centre-of-mass energy. 

At this point the events are clustered into three jets using the 
JADE algorithm~\cite{JADE}. The $y_{\rm{cut}}$ value where the 
transition from two 
to three jets occurs 
is required to be greater than 0.001. The jet with the lowest charged track
multiplicity is taken as the $\tau$ jet candidate. If at least two jets have 
the same multiplicity the $\tau$ 
jet candidate is taken to be the lowest momentum jet. The following quality 
cuts are then applied on the $\tau$ jet. It is required that the 
charged multiplicity of the $\tau$ jet be between one and three.
The angle between the $\tau$ jet and the closest 
quark jet candidate is required to be between $30^{\circ}$ and 
$125^{\circ}$. The energy of the $\tau$ jet boosted into the charged 
Higgs boson rest frame, defined as the frame recoiling 
against the hadronic system,
is required to be less than 40~GeV.

To further suppress backgrounds the following four variables are 
used:

\begin{itemize}
\item The angle $\theta_{\qq}$ between the two hadronic jets. 

\item The total transverse momentum $p_{t}^{tot}$ divided by the total 
visible energy $E_{vis}$. 

\item The production angle of the events, reconstructed from the 
sum of the momenta of the two quark jets. This production 
angle is then charge tagged with the tracks from the $\tau$ jet to form 
the variable $\theta_{prod}^{ch}$. In the 
case of two tracks in the $\tau$ jet the charge of the highest momentum 
track is used. 

\item The $\chi^2$ per number of degrees of freedom 
from a kinematic fit 
to the events using the constraints of energy and momentum conservation and 
the equality of the two masses produced in each side of the event. 
\end{itemize}

\begin{figure}[t]
\centering
  \epsfxsize=16.0cm
  \leavevmode
  \epsfbox[27 401 529 672]{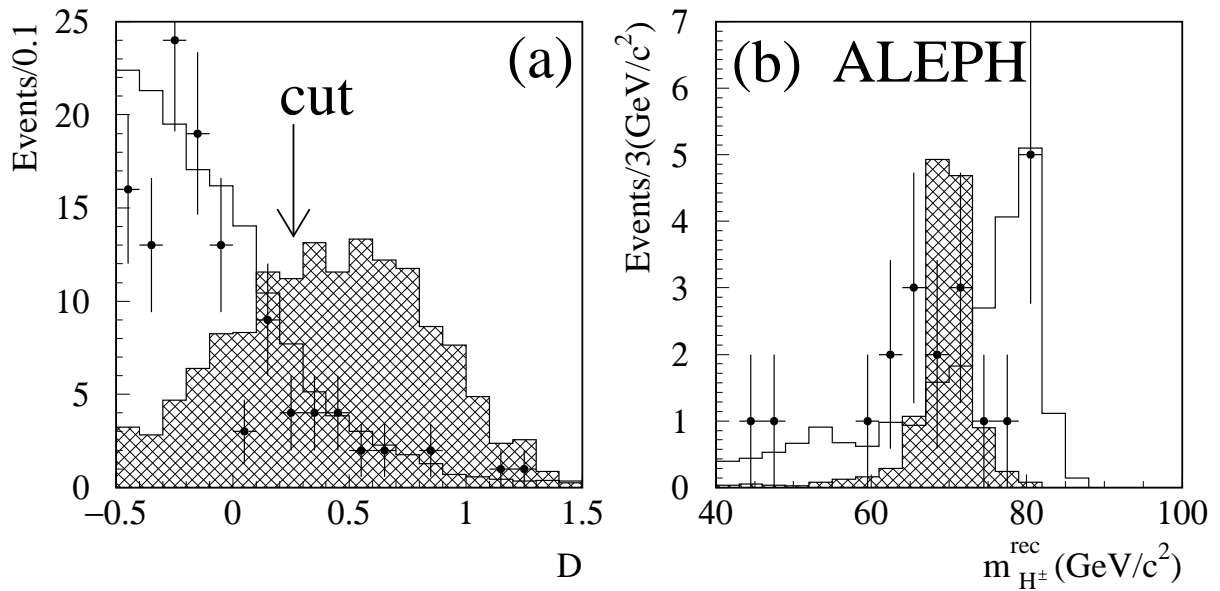}
 \caption{(a) The distribution of the discriminating variable $D$ described 
in the text for the semi leptonic channel at the level of preselection. 
(b) The reconstructed masses of the 
charged Higgs candidates after the cut on the discriminating variable. In both 
plots the points are the data, the open histograms are the 
Standard Model background and the hatched areas represent the charged 
Higgs signal expectation with $\mh$~=~70~\gevcc{}. The signal is normalised 
arbitrarily in both plots.}
 \label{fig:hphm_tncs}
\end{figure}

The four variables are linearly combined in a discriminating variable $D$, 
displayed in Fig.~\ref{fig:hphm_tncs}a.
The cut optimisation leads to $D>0.26$. 
In the data collected at $\sqrt{s}=$~189~GeV, 
20 events are selected, in agreement with 
the background expectation of~22.6. The efficiencies of selection for a range 
of masses are given in Table~\ref{effic}.
The mass of the hadronic system, and hence the mass of the charged Higgs
candidates, is rescaled such that the energy of the two jets is equal to 
the beam energy in order to improve the resolution. 
The reconstructed masses of the candidate events are displayed in 
Fig.~\ref{fig:hphm_tncs}b. The cutoff near 80~GeV/$c^2$ is due to
the influence of the $\theta_{\qq}$ variable.


The systematic uncertainty on the number of expected signal events 
is estimated to be 4.1\%. 
The main contributions are from Monte Carlo statistics (3.5\%), calibration
uncertainties (1.5\%) and uncertainty on the cross section for charged Higgs 
production (1.0\%). The systematic error on the background level is 
estimated to be 12\%. The main contributions are from Monte Carlo 
statistics (3\%), uncertainty on the cross section for the $\ww$ process 
(2\%) and from the statistics of data/Monte Carlo comparisons (10\%). 
The systematic error on the luminosity is estimated to be 0.5\%. 

\begin{boldmath}
\subsection{The \cscs{} final state}
\end{boldmath}

The hadronic decays of two charged Higgs bosons
lead to a final state of four well separated jets, which can be combined 
into two dijets with equal masses. With respect to Ref.~\cite{HCH183} the 
preselection and jet pairing method remain unchanged. 

The following five variables are used:

\begin{itemize}
\item The $\chi^2$ per degree of freedom from a five-constraint kinematic fit.
The constraints in the fit are from conservation 
of energy and momentum and the equality of the two charged Higgs masses 
in the event.
\item The production polar angle $\theta_{prod}$, {\it i.e.}, the angle 
between the charged Higgs boson momentum direction and the beam axis.
\item The difference between the largest and the smallest jet energies,
$E_{max} - E_{min}$.
\item The product of the minimum angle between any two jets, and the 
smallest jet energy, $E_{min} \times \theta_{\qq}$. 
\item The QCD matrix element ${\cal M}_{\qq}$~\cite{matele}.
\end{itemize}

\begin{figure}[t]
\begin{center}
\centering
  \epsfxsize=16.0cm
  \leavevmode
  \epsfbox[35 401 541 672]{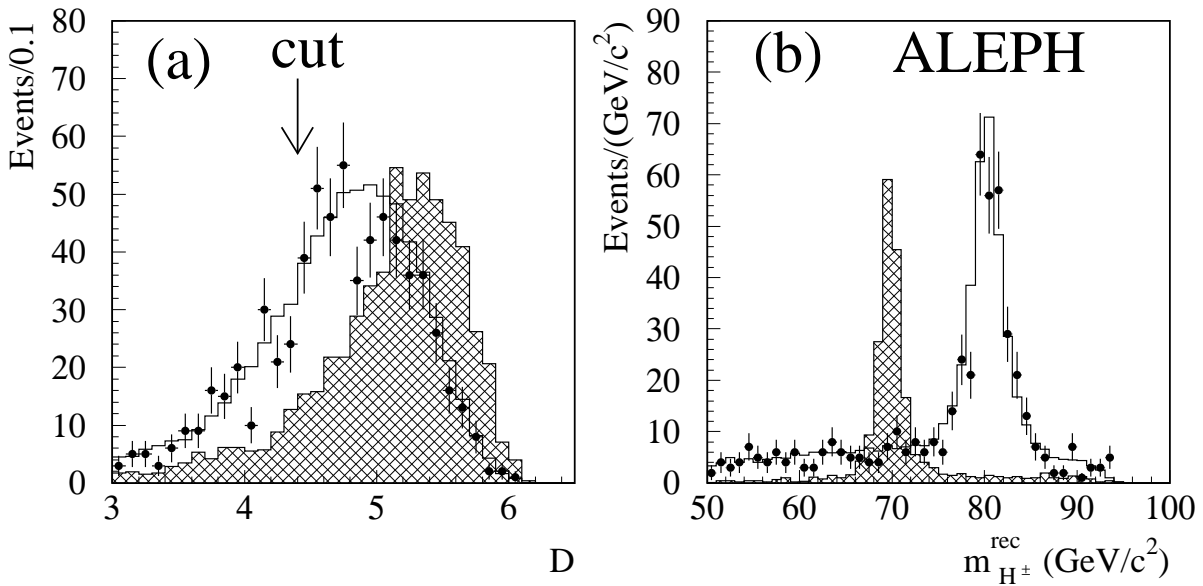}
\caption{(a) The distribution of the discriminating variable $D$
described in the 
text for the \cscs{} channel at the level of preselection. 
(b) The distribution of the Higgs candidate masses  
after the cut on the discriminating variable. In both 
plots the points are the data, the open histograms are the 
Standard Model background and the hatched areas represent the Higgs signal 
expectation with $\mh$~=~70~\gevcc{}. The signal is normalised arbitrarily 
in both plots.}
\label{fig:hphm_cscs}
\end{center}
\end{figure}

The variables are linearly combined into one discriminating variable $D$,
shown in Fig.~\ref{fig:hphm_cscs}a.
Events are accepted if $D>4.4$.
Including in $D$ the charm tag used in Ref.~\cite{HCH183} does not increase 
the discriminating power.

The analysis selects 263 events from the data for masses between 50 and
80~\gevcc{}, corresponding to a background of 294.2 events expected from 
Standard Model processes. The fitted-mass distribution of the selected 
candidates 
can be seen in Fig.~\ref{fig:hphm_cscs}b. Efficiencies are of the order of 
50\% as shown in Table~\ref{effic}.

The systematic error on the number of events expected
is estimated to be $3.0\%$. The main
contributions are from Monte Carlo statistics ($2.4\%$), statistics of 
data/Monte Carlo comparison
($1.2\%$), and knowledge of signal cross sections  
($1.0\%$). The systematic error on the background level is 
estimated to be $3\%$. The main contribution is from knowledge of 
the $\ww$ cross section ($2\%$).
The contribution of the luminosity is estimated to be 0.5\%.

\section{\label{combine}Results}

The numbers of candidates observed in the data collected at a centre-of-mass 
energy of 189~GeV are consistent with those expected from Standard Model 
processes for each of the three channels. 
Since, in addition, the mass distributions in the \cscs{} and \tncs{} 
channels do not show any significant accumulation outside the \ww{} region 
(Figs.~\ref{fig:hphm_tncs} and~\ref{fig:hphm_cscs}), the results of the three 
selections described in this letter are combined with those obtained using 
$\sqrt{s}=172$--$183~{\rm GeV}$ data to set a 95\% confidence level upper 
limit on the cross section for pair production of charged Higgs bosons. 

In setting the limits several new features with respect to 
Refs.~\cite{HCH172,HCH183} are to be noted. 
Full background subtraction is performed according to Ref.~\cite{shan}. 
The likelihood ratio test statistic is used.
The confidence levels are calculated using the semi-analytical 
approach described in Ref.~\cite{Hongbo}. 
Systematic errors are conservatively taken into account
by reducing the efficiencies and subtracted backgrounds 
by one standard deviation.
The reconstructed mass of the charged Higgs boson is used as
discriminating variable for the \cscs{} and \tncs{} channels. 

\begin{figure}[t]
\centering
  \epsfxsize=0.8\textwidth
  \leavevmode
  \epsfbox[18 142 533 677]{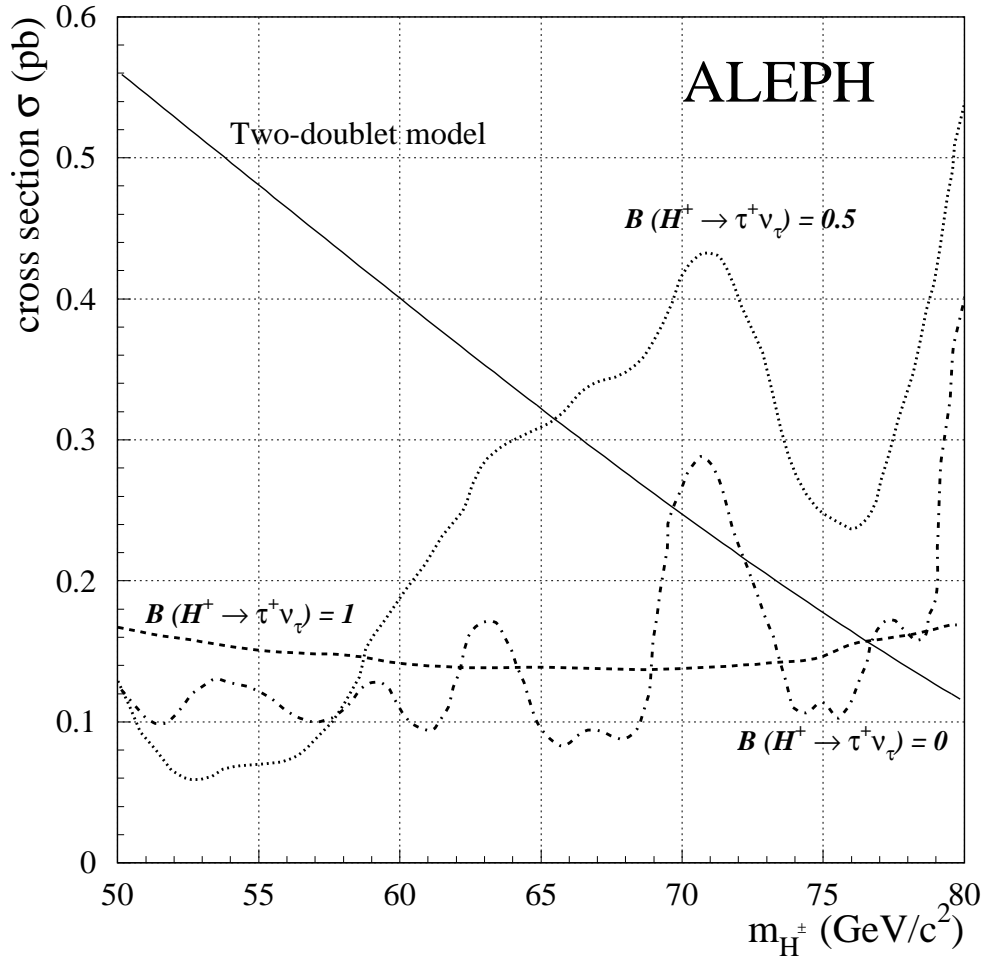}
 \caption{Upper limits at 95\% confidence level on the \hh{} production 
cross section at $\sqs = 188.6$~GeV for three values of \btn{}. 
The charged Higgs boson production cross section
is shown as a solid curve.} 
 \label{xsec_scan}
\end{figure}

The upper limit on the \hh{} production cross section at 188.6 GeV
as a function of $\mh$ is shown in~Fig.~\ref{xsec_scan} for three values 
of \btn{}. The results from lower centre-of-mass 
energies have been scaled to \sqs{} = 188.6~GeV according to the 
dependence of the cross section on the centre-of-mass energy. 

In two-Higgs-doublet models the  production cross section 
for $\hh$ depends, at lowest order, only on $\mh$. The expected cross section 
at 188.6~GeV, corrected for initial state radiation, is 
shown in~Fig.~\ref{xsec_scan}. Upper limits on the production 
cross section therefore translate into excluded domains for $\mh$.

\begin{figure}[t]
\begin{center}
\epsfxsize=0.8\textwidth
\leavevmode
\epsfbox[18 142 533 677]{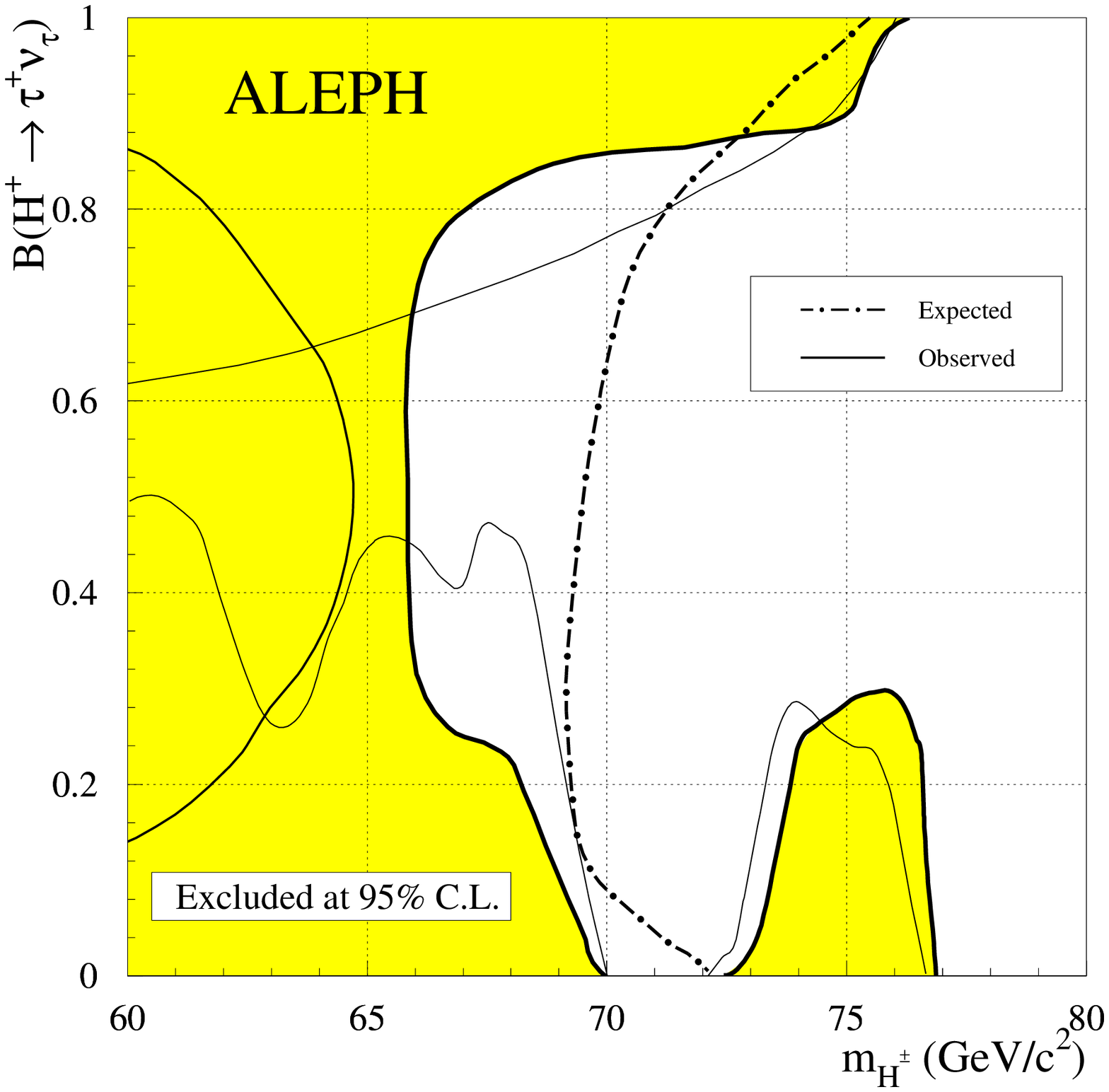}
\caption{\small Limit at $95\%$ C.L. on the mass of charged
Higgs bosons as a function
of \btn. Shown are the expected (dash-dotted) 
and observed (solid) exclusion curves for the combination of the three  
analyses, and the full 172--189~GeV data set. The shaded area is excluded at 
95\% C.L.. The observed limits from individual channels are indicated
by a thin solid line.}
 \label{combi}
\end{center}
\end{figure}

The result of the combination of the three analyses is displayed in 
Fig.~\ref{combi}.
Charged Higgs bosons with masses less than 65.4~\gevcc{} are excluded
at 95\% confidence level independently of \btn. 
The corresponding expected exclusion is 69.1~\gevcc{}. 
For the values $\btn=0$, 0.5 and 1, 95\% C.L. lower limits on $\mh$ are set 
at 69.9, 65.4 and 76.3~$\gevcc$, respectively.

\section{Conclusions}
The search for pair-produced charged Higgs bosons in the three final states
\tntn, \tncs{} and \cscs{} has been updated using 176.2~\invpb{} of data
collected at $\sqrt{s}=188.6$~GeV. No evidence of charged Higgs boson 
production has been found and upper limits have been set on the 
production cross section as a function of \btn{} and of \mh{}.
Within the framework of two-Higgs-doublet models these results exclude 
at $95\%$~confidence level 
charged Higgs bosons with masses below 65.4~\gevcc{}
 independently of \btn{} and 
assuming~$\btn+\bcs=1$. Similar results have been reported by L3~\cite{L3189}.

\vspace{0.5cm}
\noindent {\bf {\Large Acknowledgements }}
\vspace{3mm}

It is a pleasure to congratulate our colleagues from the accelerator divisions
for the successful operation of LEP at high energy.
We are indebted to the engineers and technicians in all our institutions
for their contribution to the excellent performance of ALEPH.
Those of us from non-member states wish to thank CERN for its hospitality
and support.


\begin{thebibliography}{99}



\bibitem{HCH172} ALEPH Collaboration, {\it ``Search for charged Higgs 
bosons in $e^+e^-$ collisions at centre-of-mass energies 
from 130 to 172 GeV''}, Phys. Lett. {\bf B 418}~(1998)~419.

\bibitem{HCH183} ALEPH Collaboration, {\it ``Search for charged Higgs 
bosons in $e^+e^-$ collisions at $\sqrt{s}$~=~181-184 GeV''},
Phys. Lett. {\bf B 450}~(1999)~467.

\bibitem{bib:detectorpaper} ALEPH Collaboration,
{\it ``ALEPH: a detector for electron-positron annihilations at LEP''},
Nucl. Instrum. and Methods {\bf A 294}~(1990)~121.

\bibitem{bib:performancepaper} ALEPH Collaboration,
{\it ``Performance of the ALEPH detector at LEP''},
Nucl. Instrum. and Methods. {\bf A 360}~(1995)~481.

\bibitem{WAS1} S.~Jadach, B.F.L.~Ward and Z.~W\c{a}s,
{\it ``The Monte Carlo program KORALZ, version 4.0, for the lepton or quark 
pair production at LEP/SLC energies''},
Comp.~Phys.~Commun.~{\bf 79}~(1994)~503.

\bibitem{PYTHIA} T.~Sj\"ostrand,
                 {\it ``The~PYTHIA~5.7~and~JETSET~7.4~Manual''},
                 LU-TP~95/20,~CERN-TH~7112/93,
                 Comp.~Phys.~Commun.~{\bf 82}~(1994)~74.

\bibitem{KORALW} M.~Skrzypek, S.~Jadach, W.~Placzek and Z.~W\c{a}s,
{\it ``Monte Carlo program KORALW-1.02 for W pair production at
LEP-2/NLC energies with Yennie-Frautschi-Suura exponentiation''},
Comp.~Phys.~Commun.~{\bf 94}~(1996)~216.

\bibitem{PHOT} 
J.A.M. Vermasaeren, Proceedings of the IVth International
 Workshop on Gamma Gamma Interactions, Amiens, April 1980,
 Springer Verlag (Editors G.Cochard, and P.Kessler).

\bibitem{JANOT} G.~Ganis~and~P.~Janot,
                {\it ``The~HZHA~Generator''}
                in {\it ``Physics~at~LEP2''},
                Eds. G.~Altarelli, T.~Sj\"ostrand and F.~Zwirner,
                CERN~96-01~(1996),~Vol.~2, 309.

\bibitem{slepton189} ALEPH Collaboration,
{\it ``Searches for sleptons and squarks in $e^+e^-$ collisions
at \sqs{}~=~189~GeV''},
Phys. Lett. {\bf B 469} (1999) 303.

\bibitem{JADE} JADE Collaboration, 
{\it ``Experimental investigation of the energy dependence of the 
strong coupling strength''},
Phys.~Lett.~{\bf B 213}~1988~(235).

\bibitem{matele} D.~Danckaert et al., 
{\it ``Four-jet production in $e^+e^-$ annihilation''},
Phys. Lett. {\bf 114B}~(1982)~203.

\bibitem{shan} S. Jin and P. McNamara, 
{\it ``The signal estimator limit setting method''},
physics/9812030.



\bibitem{Hongbo} H. Hu and J. Nielsen, 
{\it ``Analytic Confidence Level 
Calculations using Likelihood Ratio and Fourier Transform''}, 
physics/9906010, submitted to Nucl. Instrum. and Methods~{\bf A}.


\bibitem{L3189} L3 Collaboration, 
{\it ``Search for Charged Higgs Bosons in $e^+e^-$ Collisions at 
\sqs{}~=~189~GeV''}, Phys. Lett.~{\bf B466} (1999) 71.

\end{thebibliography}
\end{document}